\documentclass[12pt]{article}

\usepackage{graphicx}
\usepackage{amsmath,amssymb}% ,amsthm

\DeclareMathSymbol{\square}{\mathord}{AMSa}{"03}
\begin{document}

%%%\usepackage{amsmath,amssymb}% ,amsthm
%%%%%%%%%%%%%
%%%%%this command are protected don't use it in macros !!!!!!!!!!!
%%%%%%\[
%%%%%% \]
%%%%%%%%\t{a}{w}
%%%%%% \i
%%%%% \d{a}
%%%%% \P
%%%%%\b{a}
%%%%%%%%%%%%

\title{Generalized $\kappa$-Deformations and Deformed
 Relativistic Scalar Fields on Noncommutative
 Minkowski Space}

\author{Piotr Kosi\'nski\thanks{Supported by KBN
 grant 5P03B05620} \ \   and Pawe{\l} Ma\'slanka$^*$  \\
Department of Theoretical Physics,
 University of {\L}\'od\'z, \\ ul. Pomorska 149/53, 90-236, {\L}\'od\'z, Poland
 \\ \\
Jerzy Lukierski$^*$ \\
Institute for Theoretical Physics,
University of Wroc\l{}aw, \\  pl. M. Borna 9, 50-205 Wroc{\l}aw,
Poland%%%%%%%%%%%%~\thanks{Supported by Grant KBN 5PO 3B05620.}
\\ \\
 Andrzej Sitarz
%%%%%%%%%%% \footnotemark[1]
%  Andrzej Sitarz}
\\
Insitute of Physics, Jagiellonian University,
\\  ul. Reymonta 4,
 Krak\'{o}w, Poland}

%definitions
\def\f{\varphi}
\def\X{{\cal X}}
\def\M{{\cal M}}
\def\k{{\kappa}}
\def\kdef{$\kappa$-deformed }
\def\kdn{$\kappa$-deformation}
\def\poin{Poincar\'e }
\def\D{\Delta}
\def\e{\epsilon}
\def\h{\hat}
\def\ben{\begin{enumerate}}
\def\een{\end{enumerate}}
\def\gpa{\h {{\cal P}}^{(g_{\mu\nu})}}
\def\gpg{{{\cal P}}^{(g_{\mu\nu})}}
\def\U{{\cal U}}
\def\Mk{{\cal M}_\kappa}
\def\Uk{{\cal U}_\kappa}
\def\hPk{\h {{\cal P}}_\kappa}
\def\bel{\begin{equation}\label}
\def\ee{\end{equation}}
\def\tens{\otimes}
\def\r#1{(\ref{#1})}
\def\kmin{$\kappa$-Minkowski }
\def\beq{\begin{eqnarray}}
\def\eeq{\end{eqnarray}}
\def\lbl{\label}
\def\pok#1{\frac{P_0}{#1\kappa}}
\def\pokk#1{\frac{{\cal P}_0}{#1\kappa}}
\def\cop{\triangle}
\def\triv#1{#1\tens 1 + 1 \tens #1}
\def\void{\,\,\cdot\,\,}
\def\ba{\begin{array}}
\def\ea{\end{array}}
\def\Lam{\Lambda}
\def\ost{\frac1{\sqrt2}}
\def\sh{\sinh}
\newcounter{rown}
\def\bl{\setcounter{rown}{\value{equation}}
        \stepcounter{rown}\setcounter{equation}0
        \def\theequation{\thesection.\arabic{rown}\alph{equation}}
        }
\def\el{\setcounter{equation}{\value{rown}}
        \def\theequation{\thesection.\arabic{equation}}}
\def\sec{\setcounter{equation}0}
\def\p{\partial}
\def\ik{{\frac i \kappa}}
\def\t{\tilde}
\def\tf{\tilde f}
\def\pp#1#2{\frac{\p #1}{\p#2}}
\def\epk#1{e^{#1\frac {P_0}{\kappa}}}
\def\epok#1{e^{#1\frac {\partial_0}{\kappa}}}
\def\tp{\tilde p}
\def\kpoin{$\k$-\poin}
\def\<{\left<}
\def\>{\right>}
\def\ogr#1{| %\rule{0.5pt}{3ex}
_{#1}}
\def\dsp{\displaystyle}

\renewcommand\theequation{\thesection.\arabic{equation}}

%\date{}

%\maketitle

%\twocolumn
%[\maketitle
\onecolumn

\maketitle

%[maketitle
\abstract{
%\begin{abstract}
  We describe the generalized \kdn s of $D=4$ relativistic
symmetries with finite masslike deformation parameter $\k$ and an
arbitrary direction in \kdef Minkowski space being noncommutative.
The corresponding bicovariant differential calculi on \kdef
Minkowski spaces are considered.  Two
 distinguished
 cases are discussed: 5D
noncommutative differential calculus (\kdn\
in time-like or space-like direction), and 4D noncommutative
differential calculus  having the classical dimension
(noncommutative  \kdn\ in light-like direction).
 We introduce also left and
right vector fields acting on functions of noncommutative
Minkowski coordinates, and describe the noncommutative
differential realizations of \kdef \poin algebra. The \kdef
Klein-Gordon field on noncommutative Minkowski space
 with noncommutative time (standard $\kappa$-deformation)
 as well as noncommutative
null line (light-like $\kappa$-deformation)
 are discussed.
Following our earlier proposal (see \cite{koluma45,koluma2})
 we introduce an
equivalent framework replacing the local
 noncommutative  field theory by the nonlocal commutative description
 with
  suitable nonlocal star
  product multiplication rules. The modification of
 Pauli--Jordan commutator function is described and the
 $\kappa$-dependence of its light-cone behaviour  in coordinate space
  is explicitely given.
   The problem with the $\kappa$-deformed energy-momentum conservation law is
   recalled.
}
% ]

%\end{abstract}

%%%%%\maketitle

%%%%%%%%%\onecolumn

\newpage
\sec
\section{Introduction}

Recently the noncommutative framework has been studied
 in dynamical theories along the following lines:

 i) The commutative classical Minkowski coordinates $x_\mu$
  one replaces by the noncommutative ones (see e.g.
 \cite{dofrro}--\cite{rluk7})

\bel{lll1}
  [ \widehat{x}_\mu , \widehat{x}_\nu ] = i
  {\theta}_{\mu\nu}( \widehat{x} )\, .
\end{equation}
 In particular it has been extensively studied the case with
constant $\theta_{\mu\nu} (\widehat{x}) \equiv \theta_{\mu\nu}$.
In such a simple case the relativistic symmetries remain
classical, only the constant tensor $\theta_{\mu\nu}$ implies
explicite breaking of Lorentz symmetries.

ii) One can start the considerations from the generalization of
classical symmetries with commutative parameters replaced by
noncommutative ones. In order to include  in one step the
deformations of infinitesimal symmetries (Lie-algebraic
framework) and deformed global symmetries (Lie groups approach) the
Hopf-algebraic description should be used,  providing quantum
   groups which are  dual to quantum Lie algebras.

The Hopf algebra framework of quantum deformations
 \cite{koma1}--\cite{koma3} has been
extensively applied to the description of modified $D=4$
space-time symmetries in 1991--97 (see e.g. \cite{koma4}--\cite{luk2a}).
  There were
studied mostly\footnote{Since 1993 due to Majid and Woronowicz it
is known that the Drinfeld-Jimbo deformation of Lorentz algebra
with dimensionless parameter $q$ can not be extended to
$q$-deformation of Poincar\'{e} algebra without introducing
braided tensor products (see e.g. \cite{koma9}). } in some detail the
quantum deformations with mass-like parameters, in particular the
\kdef $D=4$ \poin algebra $\Uk(\h {{\cal P}})$ written in
different basis (standard \cite{koma4,koma6,koma7},
  bicrossproduct \cite{koma12} and
classical one \cite{koma18}), the  \kdef \poin group ${{\cal P}}_\kappa$
 \cite{koma11,koma12} as well
   as $D=4$ $\kappa$-deformed $AdS$ and conformal
symmetries \cite{koma22,24bis}.

 The \kdef Minkowski space $\Mk$, described by
the translation sector of the \kdef \poin group ${{\cal
P}}_\kappa$, is given by the following
 Hopf algebra \cite{koma11,koma12} \bl

\bel{1.1a} [x^{\mu},x^{\nu}]=\frac{i}{\kappa
}(\delta_{0}^{\mu}x^{\nu}- \delta_{0}^{\nu}x^{\mu})
\ee
 with
classical primitive coproduct
\bel{1.1b} \D x^\mu=x^{\mu}\tens1 +
1 \tens x^{\mu}
 \ee
  \el
 \noindent
  as well as classical antipode
($S(x^{\mu})=-x^{\mu}$) and classical counit
($\epsilon(x^{\mu})=0$). We see from \r{1.1a}-\r{1.1b} that the
space-time coordinate which is ``quantized'' by the deformation
procedure is the time coordinate $ x_{0}$, and  the nonrelativistic
$O(3)$ rotations remain unchanged. By considering different
contraction schemes there were proposed also the \kdn s along one
of the space axes,
 for example $x_3$ (this is so-called
tachonic \kdn\ \cite{luk2a} with $O(2,1)$ classical subalgebra). Other
interesting \kdn\ is the null-plane quantum \poin algebra
\cite{koma16}\footnote{We shall further call this algebra null-plane \kdef
\poin algebra. The deformation  presented in \cite{koma16} is the
particular case of generalized \kdn s of \poin algebra,
considered in \cite{koma20,koma21,luk2a}.}, with the ``quantized'' light-cone
coordinate $x_+=x_0+x_3$ and classical $E(2)$ subalgebra. Such a
\kdn\ of \poin symmetry in light-like direction has the following
two remarkable properties:
\ben
 \item[i)] The infinitesimal
deformations of null-plane \kdef algebra are described by
classical $r$-matrix satisfying the classical Yang-Baxter
equation (CYBE). As a consequence, this deformation can be
extended to larger $D=4$ conformal symmetries \cite{koma22,24bis}.
 \item[ii)] For
null-plane \kdef Minkowski space the bicovariant differential
calculus is fourdimensional, with one-forms spanned by the
standard differentials $dx_\mu$ (see \cite{koma21}) and subsequently
  the differentials
are not coboundary, similarly as in the classical case (see
Sect.\ 3). We see therefore  that the ``null-plane'' \kdn\
provides an example of 4D differential calculus on quantum
Minkowski spaces  considered by Podle\'s \cite{koma24}.
 \een

 The
generalized $\kappa$-deformation of $D=4$ \poin
 symmetries were proposed in \cite{koma20}.
 They are  obtained by introducing
   an arbitrary  symmetric Lorentzian  metric
$g^{\mu\nu}$ with signature $(+,-,-,-)$.
  Let us observe that the change of the linear basis
in standard Minkowski space with Lorentz metric tensor
$\eta^{\mu\nu}=diag(1,-1,-1,-1)$ \bel{1.2} x_\mu \to y_\mu =
R_\mu{}^\nu x_\nu \ee implies the following replacement of the
Lorentzian metric \bel{1.3} \eta^{\mu\nu} \to
g^{\mu\nu}=R^\mu{}_\rho \eta^{\rho\tau} R_\tau{}^\nu\qquad
R^\mu{}_\rho =(R_\rho{}^\mu)^T\,.
 \ee
  In \cite{koma20} there was proposed
a deformation of $D=4$ \poin group with arbitrary Lorentzian
metric $g^{\mu\nu}$. In such a case the deformed, ``quantum''
direction in standard Minkowski space is described by the
coordinate $y_0=R_0{}^\nu x_\nu$, where $R_0^\nu$ is chosen in
such a way that the relation \r{1.3} is valid. In particular one
can choose
\ben
\item[i)] For tachyonic \kdn\ (\kdef
$x_3$-direction)
\bl \bel{1.4a}
R=%\pmatrix
\begin{pmatrix} %{
 0 & 1 &
0 & 0 \cr
             1 & 0 & 0 & 0 \cr
             0 & 0 & 1 & 0 \cr
             0 & 0 & 0 & 1 %}
             \end{pmatrix}
             \quad \to \quad
g=%\pmatrix
\begin{pmatrix}
% {
 -1 & 0 & 0 & 0 \cr
             0 &1 & 0 & 0 \cr
             0 & 0 & -1 & 0 \cr
             0 & 0 & 0 & -1 % }
             \end{pmatrix}
\ee
\item[ii)]
For null-plane \kdn\ (\kdef $x_+=x_0-x_3$)
\def\ost{\frac1{\sqrt2}}
\bel{1.4b}
 \hspace{-0.5truecm}
R= %\pmatrix {
\begin{pmatrix}
 \ost & -\ost & 0 & 0 \cr
             -\ost & -\ost & 0 & 0 \cr
             0 & 0 & 1 & 0 \cr
             0 & 0 & 0 & 1 %}
\end{pmatrix}
             \ \to \
g= %\pmatrix {
\begin{pmatrix}
 0 & -1 & 0 & 0 \cr
             -1 & 0 & 0 & 0 \cr
             0 & 0 & -1 & 0 \cr
             0 & 0 & 0 & -1 %}
             \end{pmatrix}
\ee
\el
\een

The plan of our presentation is the following:

 In Sect. 2, following
 \cite{koma20} we shall present the
 \kdn\ $\Uk(\gpa)$ of the
\poin algebra $\gpg$ describing the group of motions in
space-time with arbitrary constant metric
 $g_{\mu\nu}$  with signature $(+,-,-,-)$.
  The generators
$M^{\mu\nu}$, $P_\mu$ ($P^\mu\equiv g^{\mu\nu}P_\nu$) satisfy the
following algebra:
 \bl
 % \bel{1.5a}
 \begin{eqnarray}
 \label{1.5a}
% \ba{rcl}
%\!\!\!\!\!\!\!\!\!\!\!\!\!\! \!\!\!\!
 \hspace{-3.5truecm}
 [M^{\mu\nu},M^{\rho\tau}]  &  = & i(g^{\mu\tau}
M^{\nu\rho}-g^{\nu\tau}M^{\mu\rho}
%%%%%%%% \cr
%%%%%%&&
 +
g^{\nu\rho}M^{\mu\tau}-g^{\mu\rho}M^{\nu\tau}),
% \end{array}
% \ee \bel{1.5b}
 \\ \label{1.5b}
 \cr
[M^{\mu\nu},P_{\rho}] &=&
i(\delta^{\nu}{}_{\rho}P^{\mu}-\delta^{\mu}{}_{\rho}P^{\nu}),
%\ee \bel{1.5c}
\\ \cr
\label{1.5c}
 [P_{\mu},P_{\rho}]
 &= & 0\,.
 % \end{array}
  %\ee
  \end{eqnarray}
 \el
\noindent  where $g^{\mu\nu}$ is
the symmetric metric % $g$%^{\mu\nu}$
 with the Lorentz signature. The
deformed algebra will take  the form of bicrossproduct Hopf algebra
\cite{koma7,koma25} which permits by duality the description of corresponding
\kdef quantum \poin group $\gpg$. In Sect.\ 2 we shall also
describe the deformation map and its inverse, generalizing the
results of ref.\ \cite{koma18} to the case of arbitrary metric
 $g_{\mu\nu}$ (see also  ref.\ \cite{koma37}).
 In Sect.\ 3 we shall describe the bicovariant
differential calculus on \kdef Minkowski space; in particular, it
will be explained that in the case $g_{00}=0$ the calculus has a
classical dimension and the Podle\'s condition $F=0$  selecting
the fourdimensional differential calculi in $D=4$ \cite{koma24} is
satisfied. Further, in Sect.\ 4, we shall introduce left and
right vector fields on noncommutative $\kappa $-Minkowski space
and, subsequently, write down the realizations of \kdef \poin
algebra on $\kappa$-deformed Minkowski space.
  Exploiting duality we shall
explain the relation between the realizations on noncommutative
$\kappa $-Minkowski space and known realizations (see
\cite{koma6,koma7,koma26,koma27})
  on commutative fourmomentum space.  In Sect.\ 5 we
shall describe the noncommutative plane wave decomposition of the
Klein-Gordon (KG) equation on $\kappa$-Minkowski space using
normally ordered exponentials \cite{koma12,kolumasi} for the standard
$\kappa$-deformation as well as
 for the light-cone $\kappa$-deformation.
  We shall show the relation of this approach to the
technique using nonordered exponentials, proposed by Podle\'s in
\cite{koma24}. Further, in Sect. 6, we shall discuss an equivalent
nonlocal K--G action on classical Minkowski space. In  this
 commutative framework we shall calculate   the deformation of Pauli--Jordan
function, describing $\kappa$-deformed second-quantized free KG
field and describe  the $\kappa$-deformed behaviour around the
light-cone. In Sect. 7 we present final remarks.

One should observe that recently the algebraic framework of
$\kappa$-deformed symmetries as well as some elements of
$\kappa$-deformed differential    calculus were employed for the
description of so--called doubly special relativistic (DSR)
theories (see e.g. \cite{koma28}--\cite{koma35}).
 One of the aims of this paper is to
provide some theoretical background for these more
phenomenologically - oriented  considerations.

\sec
\section{\kdef \poin Algebra $\Uk(\gpa)$ and \kdef \poin Group $\gpg_\k$ in
Arbitrary Basis}

The \kdn\ of the classical algebra \r{1.5a}-\r{1.5c} is generated
by the following $r$-matrix \cite{koma20}\footnote{In implicite form the
relation (\ref{2.1}) is present in \cite{koma15}.}
 \bel{2.1}
r=\displaystyle{\frac{i}{\kappa} }M_{0\mu}\wedge
P^{\mu}=\displaystyle{\frac{i}{\kappa}
}\,g^{\mu\nu}\,M_{0\mu}\wedge P_{\nu}\,.
 \ee
  The relations \r{1.5a} and \r{1.5c}
remain unchanged. The cross-product relation \r{1.5b} is deformed
 in the following way (we denote the \kdef generators by
$\M^{\mu\nu}=(\M^{ij},\M^{i0}$; ${{\cal P}}_{\mu}=({{\cal
P}}_{i},{{\cal P}}_{0})$; $i,j=1,2,3$):
\bl \beq \lbl{2.2a}
[\M^{ij},{{\cal P}}_{0}]&=&0\,,
\\ %\nonumber
\cr
\lbl{2.2b}
 [\M^{ij},{{\cal P}}_{k}]&=&i\k(\delta^{i}{}_{k}g^{0i}
-\delta^{i}{}_{k}g^{0j})(1-e^{-\frac{{{\cal P}}_{0}}\k})
%%%%%%%% \\
%%%%%&&
+ i(\delta^{j}{}_{k}{{\cal P}}^{i}-\delta^{i}{}_{k}{{\cal P}}^{j})\,,
\\  \nonumber
 &&
  \\
\lbl{2.2c} [\M^{i0},{{\cal P}}_{0}]
&=& i\k g^{i0}(1-e^{-\pok{}})+i{{\cal P}}^i\,,
\\
\cr
\lbl{2.2d} [\M^{i0},{{\cal P}}_{k}]
&=& -i\frac{\kappa}2 g^{00} \delta^{i}{}_k(1-e^{-2\pok{}})
%%%%%%%\cr
%%%%%%&&
-i\delta^i{}_kg^{0s}{{\cal P}}_s e^{-\pok{}}
 \cr  %%\nonumber
 && +
i g^{oi}{{\cal P}}_k(e^{-\pok{}}-1)
%%%%%%%%% \cr
%%%%%%% &&
 - \frac i {2\k}
 \delta^i{}_k {{\cal P}}_l {{\cal P}}^l
 - \frac i \kappa {{\cal P}}^i {{\cal P}}_k\,,
%%%%%%%\cr
%%%%%%%%%&& \\
\eeq
\el
where ${{\cal P}}^k\equiv g^{kl}{{\cal P}}_l$ $(k,l=1,2,3)$.
The coproducts are the following:

\bl
\beq
\lbl{2.3a}
\cop {{\cal P}}_0 & = & \triv{{{\cal P}}_0} \,,
\\ \nonumber
%\quad
 \cop {{\cal P}}_i\,&  = & \, {{\cal P}}_i
  \tens e^{-\pok{}} + 1 \tens {{\cal P}}_i\,,
%%%%%%%%%
%%%%%%\cr &&
%%%%%%
\\[6pt]
\lbl{2.3b}
\cop \M^{ij} &=& \triv{ \M^{ij}} \,,
\\
\lbl{2.3c}
\cop \M^{i0} &=&  \M^{i0}  \tens e^{-\pok{}} + 1 \tens \M^{i0}
%%%%%\cr
%%%%% &&
 + \frac1\k \M^{ij}
\tens {{\cal P}}_j\,,
%\cr
% &&
\eeq
\el
The antipodes and counits are
\bl
\beq %\nonumber
\lbl{2.4a} S({{\cal P}}_0)&=&-{{\cal P}}_0\,,
%\qquad
\cr \cr
S({{\cal P}}_i)&=& e^{-\pok{}}{{\cal P}}_i\,,
\\[10pt]
%%%%%&& \\
\lbl{2.4b} S(\M^{ij})&=&-M^{ij}\,,
%\qquad
 \cr\cr
 S(\M^{i0})
&= & -e^{-\pok{}}(\M^{i0}+\frac1\kappa \M^{ij} {{\cal P}}_j)\,,
%%%%%\nonumber
%%%%%%%%\\
\eeq
\el
\bel{2.5}
\epsilon(\M^{\mu\nu})=\epsilon({{\cal P}}^{\mu})=0\,.
\ee

The Schouten bracket $[\![\void,\void]\!]$
 (see e.g. \cite{koma25}) describing the
modification of CYBE for the classical $r$-matrix \r{2.1} is the
following \cite{koma20}
 \bel{2.6} [\![ r,r]\!] = i \frac{g_{00}}{\k^2}
\M_{\mu\nu} \wedge {{\cal P}}^\mu \wedge {{\cal P}}^\nu\,.
\ee
The relation \r{2.6} explains why the null-plane \kdn\ (see
\r{1.4b}) with $g_{00}=0$ is described by CYBE.

The formulae \r{2.2a}-\r{2.3c} describe the \kdn\ in bicrossproduct basis.

Further we would like to make the following comments:
\ben
\item[i)]
The fourmomentum Hopf algebra described by \r{1.5c} and \r{2.3a} does not
depend on the metric $g^{\mu\nu}$.
The Hopf algebra \r{1.1a}-\r{1.1b}
describing the translation sector of $\gpa$ as well as \kdef Minkowski space
is dual to the fourmomentum Hopf algebra
and is also metric-independent
\item[ii)]
The Lorentz sector of the \kdef \poin group is classical for any
metric $g_{\mu\nu}$

 \bel{2.7}
[\Lam_\mu{}^\nu,\Lam_\rho{}^\tau]=0\,,\qquad \Lam_\mu{}^\nu \
\Lam_\nu{}^\tau=\delta_\mu{}^\tau\,, \ee

\bel{2.8}
\cop(\Lam_\mu{}^\nu)=\Lam_\mu{}^\rho \tens \Lam_\rho{}^\nu\,.
\ee
The metric-dependent term occur only in the cross relation
between the Lorentz group and translation generators. Using
duality relations one obtains \cite{koma20}:
 \beq\label{2.9}
\hspace{-1truecm} [\Lam^\rho{}_\tau,a^\mu]&= &-\frac i \kappa \{
(\Lam^\rho{}_0-\delta^\rho{}_0) \Lam^\mu{}_\tau
%%%%%\cr
%%%%%%\hspace{-1truecm}
%%% &&
+ (\Lam_{0\tau} - g_{0\tau})g^{\rho\mu}\}\, .
%\nonumber
%\cr
%&& \\
\eeq
\item[iii)]
In order to introduce the physical space-time basis, with
standard Lorentz metric, one should introduce the inverse
formulae to the ones given by \r{1.2}. In such a case one obtains
e.g.\ for null-plane basis given by \r{1.4b} the light cone
coordinates usually used for the description of null plane
relativistic kinematics (see e.g.\ \cite{koma36}):
 \bel{2.10} \ba{rclrcl}
y_0&=& \ost (x_0-x_3)\, , & x_0 &=& \ost (y_0-y_3) \,,
\\[10pt]
y_3&=& - \ost (x_0+x_3)\,, & x_3 &=& -\ost(y_0+y_3) \,,
\\[10pt]
y_2&=&x_2\,, & y_1&=&x_1\,,
\ea
\ee
Similarly one can write the relation between the $\M^{\mu\nu}$ generators
 and the physical Lorentz generators $M_{\mu\nu}$:
\bel{2.11} \M^{\mu\nu} = R^\mu{}_\rho R^\nu{}_\sigma
M^{\rho\sigma}\,. \ee

\item[iv)] The mass Casimir for arbitrary metric $g^{\mu\nu}$ is
given by the following formula \cite{koma21}
 \beq \label{2.12}
\M^2(P_\mu)&=& g^{00}(2\k \sh \pok{})^2
%%%%%%%\cr
%%%%%&&
+ 4\k g^{0i} \tilde P_i
e^{\tilde %\pok2
{\frac{P_0}{2\kappa}}
} \sh \tilde %\pok2
{\frac{P_0}{2\kappa}}
%%%%%%% \cr
%%%%% &&
 + g^{rs} \tilde P_r e^{\tilde
%\pok2
{\frac{P_0}{2\kappa}}
}  P_s e^{\tilde %\pok2
{\frac{P_0}{2\kappa}} }\,. \eeq Substituting the metric \r{1.4b}
in \r{2.10} one gets the mass Casimir for null-plane $\k$-\poin
algebra, obtained firstly in \cite{koma16}.

\item[v)]
 One can extend the deformation maps, written in \cite{koma18} for standard Lorentz
metric $g^{\mu\nu}=\eta^{\mu\nu}$ and express the fourmomenta
generators ${\cal P}_\mu$, $\M_{\mu\nu}$ satisfying
\r{2.2a}-\r{2.2d} in terms of the classical \poin generators ${
P}_\mu$, $M_{\mu\nu}$ satisfying \r{1.5a}-\r{1.5c} (deformation
map) and write down the inverse formulae. We put
$\M^{\mu\nu}=M^{\mu\nu}$, and for the fourmomenta sector  the
generalization of the formulae \cite{koma18}  for general $g^{\mu\nu}$
looks as follows (see also \cite{koma37}):
 \ben
\item[a)] Deformation map
\end{enumerate}
\bl
\beq
\lbl{2.13a}
\hspace{-1truecm}
{{\cal P}}_0& = &\k\ln(\frac{P_0+C}{C-g_{00}A})\,,
\\
\lbl{2.13b}
 \hspace{-1truecm}
 {{\cal P}}_i & = &\frac{\k P_i}{P_0+C} + \frac{\k A}{P_0+C}
g_{i0}\,, \eeq \el
where
\bel{2.14} g_{00}A^2(M^2) -
2A(M^2)C(M^2)+M^2=0
 \ee
and
 \bel{2.15} M^2= g^{\mu\nu}P^\mu
P_\nu\,.
 \ee
  One can calculate that
   \beq
   \label{2.16}
\M^2& =& 4\k^2\frac{A^2}{M^2-g_{00}A^2}\,,
%\qquad
 \cr
 M^2& = &A^2(g_{00} +
\frac {4\k^2}{\M^2})\,.
\eeq
If $g_{00}\neq0$ one can put the
relation \r{2.14} in the form \bel{2.17} \tilde A^2 -
\frac1{g_{00}} C^2=M^2\,,
\ee
 where
  \bel{2.17a} \tilde A=
(g_{00})^{-\frac12} (C-g_{00}A)\,.
 \ee
 In particular if we choose
\bel{2.18n} \t A = \kappa \,,
\qquad C=g_{00}\sqrt{M^2+\k^2}
 \ee
one gets
 \beq
 \label{2.19n} \M^2 & = & \frac{2\k^2}{g_{00}}\left(-1+\sqrt{ 1 +
\frac {M^2}{\k^2}} \, \right) \,,
%\qquad
 \cr
 M^2 & = & g_{00} \M^2(1+
g_{00}\frac{\M^2}{4\k^2})\,,
\eeq
 and for $g_{00}=1$ one
obtains the formulae given in \cite{koma18}.

If $g_{00}=0$ it follows from
\r{2.14} that $A=\frac{M^2}{2C(M^2)}$; the simplest choice is provided
by
\bel{2.18}
C=\k\,,\qquad A=\frac{M^2}{2\k}\,.
\ee
In such a case the relation \r{2.16} takes the simplest possible form
\bel{2.19}
\M^2=M^2\,.
\ee
\ben
\item[b)] Inverse deformation map.
One obtains
\end{enumerate}
\bl
\beq\lbl{2.20a}
\hspace{-1truecm}
P_0&=&(C-g_{00}A)e^\pokk{} -C\,,\\[10pt]
\lbl{2.20b}
 \hspace{-1truecm}
 P_i&=&\frac{C-g_{00} A}{\k} e^{\pokk{}} {{\cal P}}_i -
g_{i0}A\,. \eeq \el In particular if $g_{00}=0$ and we choose $A$
and $C$ as in the formulae \r{2.18}, one gets the formulae \bl
\beq\lbl{2.21a}
P_0&=&\k(e^{\pokk{}} -1)\,,\\
\lbl{2.21b} P_i&=&e^{\pokk{}} {{\cal P}}_i - g_{i0}\frac{\M^2}{2\k}\,.
\eeq %\een
\el \een

\ben
\item[vi)] One can calculate the invariant volume element
\end{enumerate}
\begin{equation}\label{label2.25}
  d^4 {\cal P} = \det \left(
  \frac{\partial {\cal P}}{\partial P}\right)  \, d^4 P \, ,
\end{equation}
in the deformed fourmomentum space. For simplicity we shall put
$g_{00}=1$ and $g_{0i} =0$ in the formulae
(\ref{2.13a}--\ref{2.13b}). One gets from (2,13)
\begin{equation}\label{label2.26}
  \det \left( \frac{\partial {\cal P }}{\partial P}\right)
    = \frac{(C-A)^4}{2\kappa^4 |CA' -AC'|}
   \, e^{- \frac{3{\cal P}_0}{\kappa} }\, ,
\end{equation}
where $A' = \frac{dA}{dM^2}$ and $C' = \frac{d C}{dM^2}$. Because
the functions $A,C$ are Lorentz-invariant, the invariant measure
in deformed four-momentum space takes the form
\begin{equation}\label{2.27}
  d^4 \mu ({\cal P}) = e^{\frac{3{\cal P}_0}{\kappa}} \, d^4 {\cal
  P}\,.
\end{equation}
It should be mentioned that the formula (\ref{2.27}) can be also
derived \cite{maoe}  from the Hopf algebraic scheme without  employing
the explicite formula for the deformation map.

\sec
\section{Bicovariant Differential Calculus and Vector Fields on \kdef
Minkowski Space}

Because the $\k$-Minkowski space $\Mk$, described by the relations
\r{1.1a}-\r{1.1b}, is a unital $*$-Hopf algebra, one can define
on this noncommutative space--time the bicovariant differential calculus
 \cite{koma38,koma39,koma21}.
Following the results obtained in \cite{koma19}  Podle\'s  investigated
differential calculi on more general deformations of the
Minkowski space algebra by considering the generators $y_\mu$
satisfying the relations \cite{koma24}
 \bel{3.1} (R-1)^{\mu\nu}
{}_{\rho\tau}(y^\rho y^\tau - Z^{\mu\nu}{}_\rho y^\rho
+T^{\mu\nu})=0\,.
\ee
 The algebra \r{1.1a}-\r{1.1b} is obtained
by putting $R=\tau$ i.e.\
$R^{\mu\nu}{}_{\rho\tau}=\delta^\mu{}_\tau \delta^\nu{}_\rho$,
$T^{\mu\nu}=0$ and \bel{3.2} Z^{\mu\nu}{}_\rho =\frac i \k (
\delta^\mu{}_0 g^\nu{}_\rho - \delta^\nu{}_0 g^\mu{}_\rho)\,. \ee

 Podle\'{s} in \cite{koma24} looked for the condition restricting the algebra \r{3.1} which implies
 the existence of four-dimensional covariant
differential calculus.\footnote{In \cite{koma24} there are considered only
covariant differential calculi. It appears however that for the
case of \kdef Minkowski space \r{1.1a}-\r{1.1b} the Podle\'{s}
differential calculi are bicovariant for any metric $g_{\mu\nu}$.}
 It appears that one can
 relate the problem of existence of 4-dimensional
covariant calculus with the vanishing of a certain constant
four-tensor $F$. Using the bicovariance (with respect to the
coproduct on the Minkowski space), covariance (with respect to
the Poincar\'{e} group coaction) and the algebra comutation
relations (1.1a) we determine that in our case the Podle\'{s}
condition reads:

\begin{equation}
\label{lll3}
 F^{\mu\nu\rho}_\sigma = (\frac{i}{\kappa})^2 g_{00}
(g^{\nu\rho} \delta^\mu_\sigma - g^{\mu\rho} \delta^\nu_\sigma)
= 0\, .
\end{equation}
We see that the condition $g_{00}=0$ implies the classical
dimension $D=4$ of differential calculus; following
  \cite{koma21,koma23} one
obtains that for $g_{00}\neq0$ we get $D=5$.

Using the general construction of bicovariant $*$-calculi by
Woronowicz \cite{koma38} two cases $g_{00}\neq0$ and $g_{00}=0$ should be
considered separately. We recall that the bicovariant
noncommutative $*$-differential calculus on $\k$-Minkowski space
$\M_\k$ is obtained if we choose in the algebra of functions on
$\M_\k$ the ideal $R$ which satisfies the properties \ben
\item[i)] $R \subset \ker \varepsilon$,
\item[ii)] $R$ is ad-invariant under the action of $\Uk({{\cal P}})$
\item[iii)] $a\in R \Rightarrow S(a)^* \in R$ .
\een We do not intend to elaborate here on the Woronowicz theory;
we would like however to provide   here short   more intuitive
description.
 To make things more clear let us appeal to the
classical case. In order to define covariant calculus on Lie
group it is sufficient to consider (co-)tangent space at one
point, say the group unit. To construct vectors one can take the
Taylor expansion of any function around $e$. The derivatives are
then given by linear terms in such an expansion. The value of $f$ at $e$
is irrelevant so we can assume $f(e)=0$ this is the origin of
$ker\epsilon $ in the above definitions. Also, higher order
terms are irrelevant; this can be taken into account by considering
the ideal
 which
 consists of functions with their Taylor expansion
starting from quadratic or higher order terms (this is
counterpart of $R$ above) and dividing out by it. The Woronowicz
construction is a straightforward generalization of "classical"
procedure. The differential calculi are described by the
nontrivial generators which span $\D=\frac {\ker \epsilon }{R}$.

 It
appears that in case of $\kappa$-Minkowski space
 one can introduce the following basis in $R$
satisfying the properties i) -- iii)
 \bel{3.5} X^{\mu\nu}=y^\mu
y^\nu + \frac i \k (g^{\mu\nu} y_0 - \delta ^{\mu}{}_0 y^\nu) \,
.
\ee
 In addition, for $g_{00}\neq0$ $R=\ker \epsilon$ and in
order to obtain nontrivial ${\cal D}$ one has to reduce the basis
\r{3.5} by substracting the trace  %%(see also \cite{koma25})
\bel{3.6}
\tilde X^{\mu\nu} = X^{\mu\nu} - \frac14 g^{\mu\nu}
X^\rho{}_\rho\,.
 \ee
  We obtain two distinct cases:
   \ben
\item[a)] $g_{00}\neq0$
\\
In such a case using the kernel
$R$ with basis \r{3.6} one gets ${\cal D}$ span by
five generators $(y_0,y_i,\varphi=y^\mu y _\mu + \frac{3i}{\k} y_0)$.
Using general techniques presented in \cite{koma38}, one obtains the following
five-dimensional basis of bi-invariant forms
\bel{3.7}
\omega^\mu=dy^\mu\,,\qquad \Omega =d\varphi - 2 y_\alpha dy^\alpha\,.
\ee
The commutation relations between the one-forms and the generators
$y^\mu$ of the algebra of functions $f(y^\mu)$ an $\k$-Minkowski space
are the following
\bl
\beq\lbl{3.8a}
\hspace{-1truecm}
[dy^\mu,y^\nu]&=&\frac i \kappa g^{0\mu} dy^\nu
%%%%%%%%%\cr
%%%%%%%%%%%%%%&&
%\hspace{-1truecm}
- \frac i \kappa
g^{\mu\nu} dy^0+\frac14 g^{\mu\nu} \Omega\,,\\
\lbl{3.8b}
\hspace{-1truecm}
[\Omega,y^\mu]&=& -\frac 4 {\kappa ^2} g_{00} dy^\mu\,.
\eeq
\el
We see from the relation \r{3.8b} that the differential $dy^\mu$ is a
coboundary, i.e.
\bel{3.9n}
dy^\mu=\frac{-\k^2}{4g_{00}}[\Omega, y^\mu]
\ee
and the classical limit $\kappa\to\infty$ is singular.
The exterior products of one-forms remains classical
\beq
\label{3.9}
dy^\mu \wedge dy^\nu & =&  -dy^\nu \wedge dy^\mu\,,
%\qquad
\cr\cr
\Omega \wedge dy^\mu & = &  -dy^\mu \wedge \Omega
\eeq
and the Cartan-Maurer equation for $\Omega$ takes the form:
\bel{3.10}
d\Omega = -2 dy_\mu \wedge dy^\mu\,.
\ee
\item[b)] ${g_{00}}=0$\\
In such a case the kernel $R$ can be chosen with basis \r{3.5} and one obtains
${\cal D}$ span by four generators ($y_i$, $y_0$). One gets the fourdimensional
differential calculus and the basic  one-forms are the
differentials $dy^\mu$ satisfying the relation
\bel{3.11}
[y^\mu,dy^\nu]=\frac i \kappa (g^{\mu\nu} g_{0i} dy^i -\delta^\nu{}_0 dy^\mu)\,.
\ee
Because the one-form $\Omega$ commutes with the one-forms $dy^\mu$
(see \r{3.9n}),
the deformed calculus similarly like in classical case is not a coboundary one.

The relation \r{3.11} is the only one which is deformed --- other relations of the
differential calculus remain classical.
\een

It should be added that the left action of $\k$-\poin group ${{\cal P}}_\kappa$
 on $\k$-Minkowski
space
\bel{3.12}
\rho_L(y^\mu)=\Lam^\mu{}_\nu \tens y^\mu + a^\mu \tens I
\ee
describe the homomorphism and introduce the covariant action on the one-forms
\bel{3.13}
\ba{rcl}
\tilde \rho_l(\omega^\mu) &=& \Lam^\mu{}_\nu \tens \omega^\nu\,,
\\[10pt]
\tilde \rho_l(\Omega^\mu) &=& 1 \tens \omega^\mu\,, \ea \ee The
relations \r{3.8a}-\r{3.8b} (as well as \r{3.11}) are covariant
under the action of ${{\cal P}}_\kappa$ given by \r{3.12}-\r{3.13}. One can
introduce the right action of the $\k$-\poin group which is a
homomorphism \bel{3.14} \rho_R(y^\mu) = y^\nu \tens \t
\Lam_\nu{}^\mu - 1 \tens \t a ^\nu\t \Lam_\nu{}^\mu\,, \ee if
($\t\Lam_\mu{}^\nu, \t a^\nu$) are the generators of the quantum
\poin group $\gpg_{-\k}$, satisfying the relations of $\gpg_\k$
with changed sign of $\k$.

\sec
\section{\kdef Vector Fields and Differential Realizations
  of \kdef \poin Algebra
on $\k$-Minkowski Space $\M_\k$}

Let us introduce on $\M_\k$ a polynomial function $f(y)$ of four variables
$y_\mu$, which formally can be extended to an analytic function. The
product of generators ($y_i,y_0$) will be called normally ordered
  \cite{koma7}
if all generators
 $y_0$ stay to the left\footnote{Firstly such ordering was proposed in
  \cite{koma41}
  and applied to $\k$-\poin in \cite{koma12}; see also
   \cite{kolumasi,maoe}}. In such a way one can uniquely
relate with any  analytic function $f(y)$ on $\M_\k$ other function
$:f(y):$.

\subsection*{a) \kdef vector fields}

In the general case the differential of any function $f$ is described by five
partial derivatives. If we choose the left derivatives, we obtain
\beq
\label{4.1}
df&= &\p_\mu f dy^\mu + \p_\Omega f \cdot \Omega=
%%%%%%%%%\cr
%%%%%%%%%%%&=&
\X_\mu :f: dy^\mu + \X _\Omega :f: \Omega\,,
\eeq
In order to obtain the explicit form of the vector fields $\chi _\mu$ ,
$\chi _\Omega $ one can follow the straightforward although rather
tedious strategy. We take the differential of any normally ordered
$f$ and then use the commutation rules (1.1a) and (3.8) to get again
normally ordered expression and differentials standing to the right.
The results reads
\beq
\label{4.2}
\nonumber
%\ba{rcl}
\X_0:f: &=&-i:[\k(e^{\frac i \k \p_0}-1)
%%%%%%%%\cr
%%%%%%%%%&&
-\frac{g_{00}}{2\k} \M^2 (\frac1i\p_\mu)]f(y):\,,
%\nonumber
 \\[2mm]
\X_i:f:&=& :[e^{\frac i \k \p_0} \p_i + i g_{i0} \M^2(\frac1i \p_\mu)]f(y)]:\,,
\\
&&
\nonumber
 \\ %%%%%[2mm]
\X_\Omega:f:&=& -\frac18 \M^2 (\frac1i \p_\mu)f:\,,
%\ea
\nonumber
\eeq
where $\M^2$ is given by the formula \r{2.12}.

It is interesting to observe that the formulae \r{4.2} can be written in the
following way
\bel{4.4}
\X_\mu:f: = :P_\mu (\frac1i\p_\mu)f(y):\,,
\ee
where the relations $P_\mu(\frac1i\p_\mu)$ are obtained from \r{2.20a}-\r{2.20b}
(see also \r{2.21a}-\r{2.21b}) by substituting ${{\cal P}}_\mu=\frac1i \p_\mu$. Indeed,
in \cite{koma7} it has  been shown firstly for $g_{\mu\nu}\equiv \eta_{\mu\nu}$ that
\bel{4.5}
{{\cal P}}_\mu:f(y): = \frac1i : \frac \p  {\p y^\mu} f(y):
\ee
The relation \r{4.5} remains valid for any choice of the metric $g_{\mu\nu}$.
Note that \r{4.4} and \r{4.5} provide  the relation between
 Woronowicz \cite{koma38}
 and duality-inspired bases in the "Lie-algebra" of $\k$-\poin group, given
in \cite{koma20,koma39}.

Simpler formalism is obtained for
$g_{00}=0$, when $dx^\mu$ describes the basic one-forms.
In such a case the formula \r{4.2} can be shortened, and one obtains
\bel{4.3}
df=\p_\mu f\, dy^\mu =\X_\mu:f: dy^\mu\,.
\ee
We shall consider further the case $g_{00}=0$ and the choice of the
parameters \r{2.18} in the inverse deformation map occuring in \r{4.4}.
Because we define classical fields in normally ordered form, we shall
write down also the formulae for multiplying the normally ordered functions
by the coordinate $y^\mu$ from the left as well as from the right. One gets
\bl
 \beq
\label{4.6a}
\hspace{-0.5truecm}
y^\mu_L f(y) &\equiv & y^\mu f(y)
% \cr
%& =&
= :[y^0 \delta_0{}^\mu(1-e^{-\frac i \k \p_0})
+ y^\mu e^{-\frac i \k \p_0} ] f(y):\,,
%\eeq
%\bel{4.6b}
\\[2mm]
\label{4.6b}
\hspace{-0.5truecm}
y^\mu_R f(y)&\equiv& f(y) y^\mu
 = :(y^\mu-\frac i \k \delta^\mu{}_0
y^k\p_k)f(y): \,.
\eeq
\el

The commutation relations between the coordinates $y^\mu_L$ and
$y^\mu_R$ and left partial derivatives $\p_\mu$ are described by
two sets of formulas. Simpler relations are obtained for the
multiplicative operators $y^\mu_R$, acting on the right

\bel{4.7} [\p_\mu,y^\nu_R] = \delta_\mu{}^\nu +\frac i \k ( \p_0
\delta_\mu{}^\nu - g_{0\mu} g^{\mu\rho}\p_\rho)
\ee
% and this relation
%should be compared with the one written down in \cite{koma19}.
  For
completeness we write also the other relations \bel{4.8} \ba{rcl}
[\p_0,y^0_L]&=& 1+\frac i \k \p_0\,, \\[2mm]
[\p_0 ,y^i_L]&=& 0\,,\\[2mm]
[\p_i,y^0_L]&=& \frac i \k (\p_i - g_{0i} (1+ \ik \p_0)^{-1}
           (g^{0\mu}\p_\mu+ \frac i{2\k} g^{\mu\nu} \p_\mu\p_\nu)\,,\\[2mm]
[\p_i,y^j_L]&=& \delta_i^j -\ik g_{0i}(1+\ik \p_0)^{-1} g^{j\mu} \p_\mu \,.
\ea
\ee

Let us observe that
\ben
\item[i)] The relations \r{4.8} contain nonlocal operators $(1+\ik
\chi_0)^{-1}$, which are however well defined.
\item[ii)] One can also introduce the right partial derivatives, replacing
\r{4.1} with the following formulae:
\bel{4.9}
df=dy^\mu\tilde \p_\mu f = dy^\mu\,\tilde \chi_\mu:f:\,.
\ee
\een
Both partial derivatives $\p_\mu$, $\tilde \p_\mu$ are related by the
$*$-operation (Hermitean conjugation) in the Hopf algebra
\r{1.1a}-\r{1.1b}
\bel{4.10}
\p_\mu f = (\tilde \p_\mu f^*)^*\,.
\ee
The vector fields $\tilde \chi _\mu$ satisfy simpler relations with the
coordinates $y^\nu_L$.
 %These relations are written out in \cite{koma21}.

\subsection*{b) Differential realizations of $\k$-\poin algebra
$\Uk(\gpg)$ on \kdef
Minkowski space}

The differential realizations of \kpoin algebra have been firstly
given for standard choice of the metric
($g_{\mu\nu}=\eta_{\mu\nu}$) on commuting fourmomentum space
\cite{koma6,koma7,koma26,koma27}. In spinless case this realization can be extended to
the case of arbitrary metric $g_{\mu\nu}$ (see the formulae
\r{2.2a}\r{2.2d}) in the following way:
\bel{4.11a} {{\cal P}}_\mu \tf(p)
= p_\mu \tf(p)
\,,
\ee
 \bel{4.11b}
 \ba{rcl} \M^{ij} \tf(p) &=& i\{
\k (g^{0i}\pp{}{p_j} -
g^{0j}\pp{}{p_i})(1-\epk-)
- (g^{is} p_s\pp{}{\tp_j} - g^{js} p_s \pp{}{p_i})\}\tf(p)\,,
\ea
\ee

\beq
\label{4.11c}
\nonumber
\M^{i0} \tf(p) &=& i\{[\k g^{i0}(1-\epk-)+g^{ik}p_k] \pp{}{p_0}
-[\frac \k2 g^{00}(1- \epk{-2} )
+ g^{0s} p_s \epk-]\pp{}{p_i}
\\[2mm]
&&
+ g^{0i} (\epk- -1)p_k \pp{}{p_\k}
 + \frac1 {2\k} g^{rs} p_r p_s
\pp{}{p_i}
 \frac 1\k p_s p_k \pp{}{p_k}\}\tf(p)\,.
\eeq
{}From the basic duality relation \cite{koma7,koma26}
\bel{4.12}
\< \tf({{\cal P}}), :f(y):\> = \tf(\frac 1 i \p_\mu)f(x) \ogr{x=0}
\ee
one can derive the formula \r{4.5}
\bel{4.13}
\ba{rcl}
\<{{\cal P}}_\mu\tf({{\cal P}}),:f(y):\> &=& \frac 1 i \pp{}{y^\mu} \tf (\frac 1 i \pp{} y
)f(y) \ogr{y=0} \\[2mm]
&=& \< \tf({{\cal P}}),: \frac 1 i \pp{}{y^\mu} f(y):\>
\ea
\ee
as well as the relation valid for any  $f({{\cal P}})$
 which can be expanded in power series
\bel{4.14}
\ba{rcl}
\< \pp{\tf({{\cal P}})}{{{\cal P}}^\mu}, :f(y):\> &=& \left[ \pp{\tf}{{{\cal P}}^\mu}
\ogr{{{\cal P}}^\mu=\frac 1i \pp{}{y_\mu}} \right] f(y)\ogr{y=0} \\[2mm]
&=& \tf(\ik\pp{}{y}) i y_\mu f(y)\ogr{y=0} \\[2mm]
&=& \<\tf({{\cal P}}),:i y_\mu f(y):\>\,.
\ea
\ee
{}From the relation \r{4.14} follows that the differential realization
\r{4.11a}-\r{4.11c} on commuting fourmomentum space one can express
by making the replacements
\bel{4.15}
P_\mu \leftrightarrow \frac1i\pp{}{y^\mu} \,,\qquad \frac1i \pp{}{p^\mu}
\leftrightarrow y_\mu
\ee
as the differential realizations on the normally ordered functions on
noncommutative Minkowski space. Denoting
\bel{4.16}
\M^{\mu\nu} :f(y): = :\t \M^{\mu\nu} f(y):
\ee
one obtains the formula \r{4.3} as well as
\bl
\beq
\label{4.17a}
M^{ij} :f(y): &=& :[\k(1-\epok{-i})(g^{0i}y^j
-g^{0j}y^i)
 i(g^{is} y^j - g^{js} y^i)\pp{}{y^s}]f(y):\,,
\\[2mm]
\label{4.17b}
 M^{i0} :f(y): &=& :[-i y^0 (i\k g^{i0}(1-
\epok{-i})
 -
  g^{ik}\pp{}{x^\mu} %\\[2mm] &&
   \frac \k2 g^{00}y^i(1- e^{\frac{ -2\p_0}{\k}})
\nonumber
   \\[2mm]
  &&
    -ix^i g^{0k}
          \pp{}{y^k}\epok{-i}
+ ig^{0i}(\epok{-i}-1) y^k\pp{}{y^k}
  \cr
  &&
  -
  \frac1{2\k} y^i g^{rs} \frac {\p^2}{\p y^r\p y^s}
+\frac 1\k g^{is} y^r \frac {\p^2}{\p y^r \p y ^s} f(y):\,.
\eeq
\el

The relations between the functions $:~\f(x)~:$ and $\tf(p)$ carrying
respectively the realizations \r{4.17a}-\r{4.17b} and
\r{4.11b}-\r{4.11c} can be derived from the following normally ordered
Fourier transform
\bel{4.18}
:f(y): = \int d^4p\, \tf(p): e^{-i p_\mu y^\mu} :\,,
\ee
where $y^\mu$ satisfies the relations \r{1.1a}-\r{1.1b}.

The realization \r{4.11b}-\r{4.11c} can be simplified if we introduce
the nonlinear Fourier transform

\bel{4.19}
:f(x):= \int d^4 q\, \t F(q) : e^{-i{{\cal P}}_\mu(q) y^\mu}:\,,
\ee
where ${{\cal P}}_\mu(q)$ is given by the relations \r{2.13a}-\r{2.13b}.
Because it can be checked that the nonlinearities in \r{4.11b}-\r{4.11c}
are described as follows:
\bel{4.20}
\M^{\mu\nu} ({{\cal P}}, \pp{}{{{\cal P}}}) = q^\mu({{\cal P}}) \pp{{{\cal P}}_\rho}{q_\nu({{\cal P}})}
\pp{}{{{\cal P}}_\rho}- q^\nu({{\cal P}}) \pp{{{\cal P}}_\rho}{q_\mu({{\cal P}})}
\pp{}{{{\cal P}}_\rho}\,,
\ee
provided  $q^\mu({{\cal P}})$ describes the inverse deformation map (see
\r{2.20a}-\r{2.20b}), one obtains
\bel{4.21}
\M^{\mu\nu}:f(x): = \int d^4 q \t \M^{\mu\nu} \t F (q) : e ^{- i
{{\cal P}}_n(q) y}:\,,
\ee
where we get classical Loentz algebra realization
\bel{4.22}
\t\M^{\mu\nu} = \frac 1 i (q^\mu \pp{}{q_\nu} - q^\nu \pp{}{q_\mu})\,.
\ee
We see therefore that the nonlinear Fourier transform \r{4.19} relates
the $\k$-covariant functions on noncommutative \kmin space with the
functions on commutative classical fourmomentum space transforming
under classical relativistic symmetries.

\sec
\section{\kdef Klein-Gordon Fields on \kdef Noncommutative Minkowski Space}

\subsection*{a) $\kappa$-Deformed Minkowski Space and Fifth Dimension}

Let us consider the scalar field $\Phi(y)$ on the noncommutative
\kmin space \r{1.2} as the following normally ordered Fourier
transform \bel{5.1} \Phi(y)=\frac1{(2\pi)^4}\int d^4p \t\Phi(p) :
e^{ipy}:\,,
 \ee where we recall that
$py\stackrel{df}{=} p_\mu g^{\mu\nu}y_\nu = p_\mu y^\mu$
and
\bel{5.2} :e^{ipy}: = e^{p_0 x^0} e^{i p_i x^i}\,.
\ee
 The
$\k$-invariant wave operator is given by the formulae
 \beq
 \label{5.3}
g^{\mu\nu} \p_\mu\p_\nu \Phi(y)& =& \frac1{(2\pi)^4} \int d^4 p \t
\Phi(p)
 : g^{\mu\nu} \chi _\mu\chi_\nu e^{ipy} :\,,
\eeq
 where the nonpolynomial vector fields are given by the formulae
\r{4.1}-\r{4.2}. For any $g_{00}\neq 0$ one gets using the
relation \r{2.16}
\bel{5.4} :g^{\mu\nu} \chi_\mu\chi_\nu e^{ipy}:
= A^2 (g_{00} + \frac {4\k^2}{\M^2 (p_\mu)} ) : e^{ipy}:\,,
\ee
where $\M^2$ is given by \r{2.12}. For the special choice
\r{2.18n}, following the
relation \r{2.19n}  %%!!!!!!
  one obtains
  \bel{5.5} g^{\mu\nu}\p_\mu\p_\nu\Phi(y)+
\frac{g_{00}^2}{4\k^2}(\M^2)^2 \Phi(y) = -\M^2 \Phi(y) \,,
\ee
Introducing the fifth derivative $\p_\Omega \equiv \p_4$ (see
\r{4.1}) and five-dimensional metric tensor ($A,B=0,1,2,3,4$)

\bel{5.6} g^{AB}=
% \pmatrix{
\begin{pmatrix}
 g^{\mu\nu} & 0 \cr
 0 & g^{44} %}
 \end{pmatrix}
 \,,\qquad g^{44} =\frac{16 g^{00}}{\k^2}
 \ee
one can write the relation \r{5.5}the following five-dimensional
Klein-Gordon equation (see also~\cite{kolumasi})

\bel{5.7} g^{AB}\p_A\p_B \Phi(y) = -\M^2 \Phi(y)\,.
%\label{klm5.7}
\ee
 We obtain
therefore the result that if $g_{00}\neq0$ the deformed mass
Casimir is described by the five-dimensional noncommutative wave
operator, what is linked with the five dimensions of differential
 calculus.

 In principle one can generalize the formulae
 (\ref{5.1}--\ref{5.7})for the Fourier transform with any
 ordering of the noncommutative coordinates $y_\mu$. In particular
 following Podle\'{s} \cite{koma24} it is interesting to describe
the action on the noncommutative d'Alambertian (see \r{5.3}) on
nonordered exponentials $e^{ipy}$. In such a case the Fourier
transform \r{5.1} is nonordered \bel{5.8} \Phi(y)= \frac
1{(2\pi)^4} \int d^4 \t p \t{\t \Phi}(\tp) e^{i\t py} \,. \ee
 In
order to compare the deformed mass shell conditions satisfied by
the Fourier transforms $\t\Phi(p)$ (see \r{5.1}) and $\t{\t \Phi}(p)$ %%%%%%!!!!!
we observe that (see also \cite{kolumasi})
 \bel{5.9} e^{i\t p_0 y_0 - \vec{\t p} \vec
y} = e^{ip_0 y^0} e^{-i\vec p \vec y} = : e^{ipy}: \, \ee
where, due to
 the relation \r{1.1a} , \bl
\bel{5.10a} p_i=\frac \kappa {\t p_0} (1 - e^{-\frac{p_0}\k})\t
p_i\,,\qquad p_0=\t p_0 ,\ee
or inversely
\bel{5.10n} \t p_i = -
\frac {p_0}{2\k \sin \frac {p_0}\k} (1+ e ^{-\frac {p_0}\k}
)p_i\,,\qquad \t p _0 =p_0\,. \ee \el For the case of standard
\kdn\ ($q^{\mu\nu}=\eta^{\mu\nu}$) one can show that \bel{5.11}
\ba{rcl} \M^2(p_\mu) &=& (2\k\sinh\frac{p_0}{2\k})^2 - \vec p^2
e^{-\frac
{p_0}\k}
\\[2mm]
&=& \frac {4\k^2}{p_0^2} (\sinh \frac {\t p_0}{2\k})^2 (\t p_0^2
- \vec {\t p}^2)\,. \ea \ee
 The rhs describes the mass-shell
condition obtained by Podle\'s in \cite{koma24}
   derived however under the
assumption that the fourdimensional differential calculus on
quantum Minkowski space does exist.

%%%%%%%

It should be mentioned that the use of the Fourier decomposition
of \kdef free KG field on normally ordered exponentials has the
advantage of reproducing \kdef fourmomentum composition law
$p_\mu'' = \Delta^{(2)}_\mu(p,p')$
\bel{5.17} \ba{rcl}
p_0''&=&p_0 +p_0'\,,\\[10pt]
p_i''&=&p_i e^{-\frac{p_0}{\k}} +p_i'\,, \ea
\ee
which is
described by the coproduct relations for \kdef \poin algebra (see
(\ref{2.3a})). Indeed, one obtains
 \bel{5.18} :e^{-p_\mu y^\mu}::
e^{-p_\mu' y^\mu}: = : e^{-i p_\mu''y^\mu}:\,. \ee
 The $n$-fold
product of normally ordered exponentials leads to the formula
 \beq
 \label{5.19} && :e^{-i
p_\mu^{(1)} y^\mu} : \ldots : e^{-i p_\mu^{(n)} y^\mu}
: = : e^{-i
\D_\mu^{(n)} (p_{\mu}^{(1)} \cdots p_\mu^{(n)} )y^\mu}:\,,
\eeq
where
\bl
\bel{5.20a} \D_0^{(n)} (p_{\mu}^{(1)} \cdots
p_\mu^{(n)} ) = \sum_{k=1}^n p_0^{(k)} \ee \bel{5.20b} \D_i^{(n)}
(p_{\mu}^{(1)} \cdots p_\mu^{(n)} ) = \sum_{k=1}^n p_i^{(k)} \exp
\frac 1 \k \sum _{l=k+1}^n p_0^{(l)} \,. \ee \el

The formulae \r{5.19}-\r{5.20b} are important if we wish to
construct the local \kpoin-covariant vertices, by introducing the
local polynomials of the field $\Phi(y)$.

%%%%%%%

\subsection*{b) $\kappa$-deformed Klein-Gordon field
 induced by the  light cone $\kappa$-deformation}

{}From the proportionality of $g^{44}$ to $g_{00}$ it follows that
if $g_{00}=0$ the equation \r{5.7} contains only the
fourdimensional noncommutative d'Alembert operator, i.e.\ it
takes the form
 \bel{5.12} (g^{\mu\nu}\p_\mu\p_\nu+m^2)\Phi(y) =0
\ee
 in
accordance with the dimension four of differential
 calculus and  the relation \r{2.19}, which takes the form (we
recall that $g_{00}=0$):
 \bel{5.13} :g^{\mu\nu}\chi_\mu\chi_\nu
e^{ipy}:= -\M^2 (p_\mu) : e^{ipy}:\,, \ee
 where for the choice of
null-plane \kdn\ $g_{i0}=\delta_{i3}$ one gets ($r,s=1,2$)
\bel{5.14}
\M^2(p_\mu)=4\k p_3 e^{\frac{p_0}{2\k}} \sinh{\frac{p_0}{2\k}
+g^{rs}p_r p_s e^{\frac{p_0}\k}}\, . \ee
 In such a case  the solution of free KG
field \r{5.12} can be written as follows:
\bel{5.15} \ba{rcl}
\Phi(y)&=&\dsp \frac 1{(2\pi)^4} \int d^4p \, \delta (4\k p_3
e^{\frac{p_0}{2\k}}\sinh
\frac{p_0}{2\k}
\dsp + g^{rs}p_r p_s e^{\frac{p_0}{\k}} - m^2 )a (p) : e^{ipy}:
\\[2mm]
&=&
\dsp \frac1{(2\pi)^4} \int \frac{ d^2 p\, dp_0}{4\k \sinh
\frac{p_0}{2\k}}
 e^{-\frac {p_0}{2\k}}
  a(p_1,p_2,p_0)
:e^{i(p_r y^r + \omega_3 y^0 +p_0 y^3)}:\,, \ea \ee
 where
\bel{5.17b} \omega_3(p_1,p_2,p_0) = \frac { e^{-\frac
{p_0}{2\k}}}{4\k\sinh \frac{p_0}{2\k}} (\vec p^2
e^{\frac{p_0}{2\k}} - m^2)\,. \ee
 The relation \r{5.17} describes
the mass-shell condition for \kdef null-plane dynamics.
 If $m=0$ one gets from (\ref{5.17b})

\bel{5.21} \omega_3^{m=0}(p_1,p_2,p_0) =
  \frac
{\overrightarrow{p}^2 }{4\k\sinh \frac{p_0}{2\k}}
\,
%%%%%%%%%
 \mathop{\longrightarrow}\limits_{\kappa\to \infty}\,
 \frac{\overrightarrow{p}^2 }{2p_0}
\,, \ee
in accordance with the light-cone
 quantization kinematics \cite{koma36}

\sec
\section{$\kappa$-Deformed Klein-Gordon Fields on Commutative Space--Time
and $\kappa$-Deformed Pauli--Jordan Commutator Function}

{\bf a) Noncommutative action and integration over $\kappa$-Minkowski space.}

In order to describe the noncommutative action and derive the field
equations (\ref{5.5}) or (\ref{5.7}) from action principle it is
sufficient to define the integral of the ordered exponential (\ref{5.9})
over $\kappa$-deformed Minkowski space.
Following \cite{kolumasi,koluma45} we postulate
\begin{equation}\label{bel6.1}
  \frac{1}{(2\pi)^4} \, \iint d^4y\, : e^{ipy} : = \delta^4 (p) \, .
\end{equation}
{}From (\ref{bel6.1}) and (\ref{5.1}) follows that
\begin{equation}\label{bel6.2}
  \iint d^4 y \, \Phi(y) = \int d^4 \, p \, \delta^4 (p) \widetilde{\Phi}(p) =
  \widetilde{\Phi}(0)\, .
\end{equation}
Further using (\ref{5.13})
one gets
\begin{eqnarray}\label{bel6.3}
\hspace{-0.5truecm}
  \iint  d^4y \, \Phi_1(y) \Phi_2(y)
&  = &
  \int d^4   p_1 \int d^4 p_2 \Phi_1(p_1)\,  \Phi_2 (p_2)
\delta(\Delta^{(2)} (p_1, p_2))
  \cr
&  = &
  \int d^4 p \Phi_1 (\overrightarrow{p_1} p_0)\, \Phi_2
  (- \overrightarrow{p} e^{\frac{p_0}{\kappa}}, - p_0) \, ,
\end{eqnarray}
i.e. we obtain $\kappa$-deformed convolution formula.

The formula (\ref{bel6.1}) is invariant under the Poincar\'{e} transformation
of the noncommutative $\kappa$-Minkowski coordinates, described by the
formulae (\ref{3.13})  (left action of the $\kappa$-Poincar\'{e} group)
or (\ref{3.14}) (right action of the $\kappa$-Poincar\'{e} group).
Choosing the formulae (\ref{3.14}) one should show that
\begin{equation}\label{bel6.4}
  \iint d^4y \, e^{i(p_\kappa y^\kappa \otimes \Lambda_\kappa^{\ \nu}
  - p_\kappa \otimes a^\kappa \Lambda_\kappa^{\ \nu}})
  = \delta^4(p) \otimes 1 \, ,
\end{equation}
where ($a_\mu, \Lambda_\kappa^{\ \nu}$) describe the noncommutative
parameters of $\kappa$-deformed Poincar\'{e} group
 \cite{koma11,koma12,koma20}.
Indeed, it can be shown after nontrivial calculations
  (see \cite{koluma45},  Appendix)
    that the formula (\ref{bel6.4}) is valid.

We propose the noncommutative KG action in the following form
\begin{equation}\label{bel6.5}
  S = \frac{1}{2} \iint d^4y \, \Phi^+ (y) ( \widehat{\square} + m^2) \Phi(y)\, ,
\end{equation}
where $\widehat{\square}= \widehat{\partial}_\mu \widehat{\partial}^\mu$ and
 from the formula
\begin{equation}\label{bel6.6}
  \Phi^+ (y) = \frac{1}{(2\pi)^4} \int d^4p \, \widetilde{\Phi}^+ (p) :
  e^{ip\widehat{x}}:
\end{equation}
it follows that
\begin{equation}\label{bel6.7}
  \widetilde{\Phi}^+ (\overrightarrow{p},p_0 ) =
  e^{- \frac{3p_0}{\kappa} } \widetilde{\Phi}^* (- e^{\frac{p_0}{\kappa}}\, \overrightarrow{p}, -p_0)\,.
\end{equation}

{\bf b) Nonlocal commutative $\kappa$-deformed Klein-Gordon theory
 and $\kappa$-deformed star product multiplication.}

The $\kappa$-deformation implies the noncommutative $\kappa$-Minkowski
 space with noncommutative coordinates $y_\mu$ and  commutative fourmomentum
  space (coordinates $p_\mu$). From the Fourier transform
   $\widetilde{\Phi}(p)$ (see (\ref{5.1}) one can obtain also
   a standard relativistic field $\phi(x)$ on classical Minkowski
   space with coordinates $x_\mu$, by performing classical Fourier
   transform
\begin{equation}\label{bel6.8}
  \phi(x) = \frac{1}{(2\pi)^4}\int d^4p \,
\widetilde{\Phi}(p)\, e^{ipx}\, ,
\end{equation}
where for simplicity we employ  in forumomentum space the
standard integration measure (the $\kappa$-invariant one is given
by the formula (\ref{2.27})). In the limit $\kappa \to \infty$ the
noncommutative Fourier transform (\ref{5.1}) and classical one given by
(\ref{bel6.8}) coincide.

The multiplication of two field operators $\Phi_1(y), \Phi_2(y)$
is translated into homomorphic multiplication of their classical
counterparts $\Phi_1(y),\Phi_2(y)$ if we postulate the following
star multiplication of classical Fourier exponentials (see
(\ref{5.17}--\ref{5.18})
\begin{equation}\label{bel6.9}
  e^{ipx}\ast e^{ip'x} = e^{i\Delta^{(2)}(p,p')x}\, .
\end{equation}
We obtain
\begin{eqnarray}\label{bel6.10}
&&  \Phi_1(y)\Phi_2(y)  \Longleftrightarrow  \phi_1(x)\ast
\phi_2(x)
%%%%%%%%%  \cr
%%%%%%%%%%%  &&
  = \frac{1}{(2\pi)^4} \int d^4p \, d^4p'\, \Phi_1(p)
  \Phi_2(p')
%%%%%  \cr
%%%%%%  && \cdot
 e^{i\Delta^{(2)}(p,p')x}
\end{eqnarray}
and one gets (compare with (\ref{bel6.3})
\begin{equation}\label{bel6.11}
  \iint d^4y \, \phi_1(y)\phi_2(y) =
  \int d^4x \phi_1(x)\ast \phi_2(x)\, .
\end{equation}
Similarly using (\ref{bel6.7}) one gets
\begin{eqnarray}\label{bel6.12}
  \iint d^4 y   \, \Phi_1(y)^+ \Phi_2(y)
%%%   \cr
%%%%%&&
 & = &  \int d^4 x
[exp - i \frac{3\partial_t}{\kappa}
  \phi^\ast_1(x)] \ast \phi_2(x)
  \\
  & = &   \int d^3 x \, dx_0 \phi^\ast_1 ( \overrightarrow{x}, x_0
  - \frac{3i}{\kappa})   \ast \phi_2( \overrightarrow{x},x_0) \, .
  \nonumber
\end{eqnarray}
We see that the local multiplication of the fields on
noncommutative Minkowski space is replaced by nonlocal
star-multiplication on classical Minkowski space, in accordance
with the diagram depicted on Fig. 1 \cite{koluma45}:
\begin{figure}[h]
\begin{center}
\includegraphics[width=12cm,angle=0]{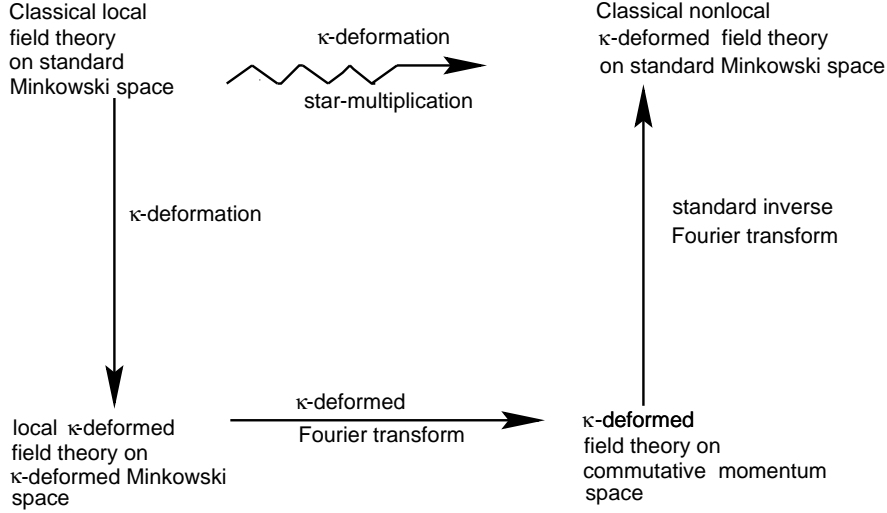}
\end{center}
\caption{Relation between noncommutative and commutative
$\kappa$-deformed field theories.}
 \label{diag}
\end{figure}

{\bf c) $\kappa$-deformed scalar Green functions: commutator function
 and the propagator. }

 The $\kappa$-deformed commutator function (Pauli--Jordan
 function) is given by the formula (we assume $\kappa > 0$):

\begin{eqnarray}\label{6.13}
  \Delta_\kappa (x)& \equiv& \frac{1}{(2\pi)^4} \int d^4p \,
  \varepsilon(p_0) \delta(M^2(p)
    (1-
  \frac{M^2(p)}{4\kappa^2})-m^2)
  e^{-ipx}
  \cr\cr\cr
   &\equiv & \Delta^+ _\kappa (x) -\Delta^- _\kappa (x)
\end{eqnarray}
where
\begin{eqnarray}\label{6.14}
  \Delta_\kappa^{\pm} (x) & \equiv & \frac{1}{(2\pi)^4} \int d^4p \,
  \theta(\pm p_0)
  \delta(M^2(p)
   (1- \frac{M^2(p)}{4\kappa^2})-m^2)e^{-ipx}\, .
   \cr &&
\end{eqnarray}
We get
\begin{eqnarray}\label{6.15}
\delta(M^2(p) (1- \frac{M^2(p)}{4\kappa^2})-m^2)
&= &  \delta(- \frac{1}{4\kappa^2}(  M^2(p)-m^2_+)
  (M^2(p) -m^2_-))
\cr
&  = &
  \frac{1}{\sqrt{1+m^2/\kappa^2}} \, (\delta (M^2(p) -m^2_+)
 +
  \delta(M^2(p) -m^2_-))
\cr
&  = &
  \frac{1}{\sqrt{1+m^2/\kappa^2}} \, \sum\limits_i \frac{1}{\left|\frac{dM^2(p)}{dp^0}\right| }
  \delta( p^0 - p^{0(i)})\, ,
\end{eqnarray}
\noindent where we summ  over all solution for $p_0$
satisfying the  following
 $\kappa$-deformed mass-shell condition

\begin{equation}\label{6.16}
  M^2(p) (1 -   \frac{M^2(p)}{4\kappa^2} )   -m^2 =0\,.
\end{equation}
We obtain
\begin{eqnarray}\label{6.17}
  \Delta^{\pm}_\kappa (x)
& \equiv &
 \frac{1}{(2\pi)^4}
  \sum\limits_i \int \frac{d^3\overrightarrow{p} \, e^{-ipx} \theta(\pm  p_0)}
  {\sqrt{1+ \frac{m^2}{\kappa^2} }\, | \frac{dM^2(p)}{dp_0}
  |}{\Bigg|}_{p_0= p_0^{(i)}}
%%%%%%%%%%  \nonumber
\end{eqnarray}
Further from $M^2(p) -m^2_{\pm} =0$ it follows that
\begin{equation}\label{6.18}
  e^{\frac{P_0}{\kappa}} ( \overrightarrow{p}^2 - \kappa^2 )
  - \kappa^2\,  e^{-\frac{P_0}{\kappa}} \mp \sqrt{1 +
  \frac{m^2}{\kappa^2}}= 0\, .
\end{equation}
Let us check that  $ \frac{dM^2(p)}{dp^0}\neq0$. Indeed, if
 $ \frac{dM^2(p)}{dp^0} = 0$, then
\begin{equation}\label{6.19}
  \frac{1}{\kappa} (  e^{\frac{P_0}{\kappa}} (
  \overrightarrow{p}^2 - \kappa^2 )
  +\kappa^2   \,  e^{-\frac{P_0}{\kappa}} ) = 0\,.
\end{equation}
{}From (\ref{6.18}--\ref{6.19}) one derives that
\begin{equation}\label{6.20}
  2\kappa^2 \left(  e^{-\frac{P_0}{\kappa}}   \pm
\sqrt{1      + \frac{m^2}{\kappa^2} }
   \right) = 0 \, .
\end{equation}
For real $p_0$  the sgn $"+"$ should be discarded. Then we obtain
\begin{equation}\label{6.21}
   e^{-\frac{P_0}{\kappa}} = \sqrt{1      + \frac{m^2}{\kappa^2}}
\end{equation}
and from (\ref{6.19})it follows that
\begin{equation}\label{6.22}
  \overrightarrow{p}^2 + m^2 = 0\, .
\end{equation}
%%%%%%%%%%%%%%%%!!!!!!!!!!!!!!!

%%%%%%%%%%%%%%!!!!!!!!!
which is impossible.
Because $ \frac{dM^2(p)}{dp_0}$ does not vanish, the only short
distance ($x^\mu \sim 0$) singularities of $\Delta^{\pm}_\kappa(x)$
can be generated by divergent integral over $\overrightarrow{p}$.
Let us observe however that the on-shell values of
$e^{\frac{P_0}{\kappa}}$ behave as
$\frac{1}{|\overrightarrow{p}|}$ for $\overrightarrow{p} \to
\infty$. For  $\Delta^{(+)}_\kappa$ the condition $p_0 > 0$
($\kappa >0$!) for large $\overrightarrow{p}$ is not valid, the
integration over $|\overrightarrow{p}|$ is truncated, and therefore short
distance singularities do not occur. For $\Delta^{(-)}_\kappa$
the situation is different -- we have two real solutions of
$\kappa$-deformed mass-shell condition (for $m_+$ and $m_-$)
with       negative $p_0$

\begin{equation}\label{6.23}
e^{\frac{P_0}{\kappa}} \sim \frac{\kappa}{|\overrightarrow{p}|}\,
, \quad  p_0 = - \kappa \ln (\frac{|\overrightarrow{p}
|}{\kappa})\, ,
\end{equation}
then

\begin{eqnarray}\label{6.24bis}
&& \left(\frac{dM^2(p)}{dp_0}\Bigg|_{p_0  =  p_0^{(i)}}
  \sim 2 | \overrightarrow{p}|
 \right)\, .
%%%%%&&
\end{eqnarray}
We get
\begin{eqnarray}
  \Delta^{(-)}_\kappa (x)
&=&
  \frac{1}{(2\pi)^4 \sqrt{1+ \frac{m^2}{\kappa^2} }}
  \int\limits^\infty_0
  \frac{d^3|\overrightarrow{p} |}{|\overrightarrow{p} |}
  \, e^{i \overrightarrow{p}\overrightarrow{x}} \,
  e^{i\kappa \ln (\frac{|\overrightarrow{p} |}{\kappa} )x^0 }
\cr\cr \cr
%%%% &&
& & =  \frac{2}{(2\pi)^3 \sqrt{1+ \frac{m^2}{\kappa^2} }}
  \int\limits^\infty_0
  \frac{d|\overrightarrow{p} |}{|\overrightarrow{x} |}
  \sin (|\overrightarrow{p} |\overrightarrow{x} | )
 (\frac{|\overrightarrow{p} |}
  {\kappa} )^{i\kappa x^0}
\cr\cr  \cr
&& =
\frac{-i}{(2\pi)^2 |\overrightarrow{x} |\sqrt{1+
\frac{m^2}{\kappa^2} }}
 \left( \int\limits^\infty _0 dp\,
e^{ip|\overrightarrow{x}|}
(\frac{p}{\kappa} )^{i\kappa x^0}
- \int\limits^\infty_0
dp \, e^{ip|\overrightarrow{x}|}
(\frac{p}{\kappa} )^{i\kappa x^0}
 \right)
\cr\cr \cr
&& =
\frac{-i\kappa}{(2\pi)^2 \, |\overrightarrow{x} | \sqrt{1+
\frac{m^2}{\kappa^2} }}
\left(
(-i \kappa | \overrightarrow{x}|)^{-1 -i\kappa x^0}\,
\Gamma(1 + i \kappa x^0 )
  - ( i \kappa |\overrightarrow{x} |)^{1-i\kappa x^0}
\, \Gamma (1 + i \kappa x^0 )
 \right)
 \nonumber
 \\[10pt]
&&
=
\frac{-i \kappa \, \Gamma (1 + i \kappa x^0) 2i
 \cosh ( \frac{\pi}{2}\kappa x^0)}
 {(2\pi)^2 \, |\overrightarrow{x} | \sqrt{1+
\frac{m^2}{\kappa^2} }(\kappa | \overrightarrow{x}|)^{1+ i\kappa x^0}}\, .
\cr
&&
\end{eqnarray}

Using (\ref{6.13}) we obtain finally that
\begin{equation}\label{6.24}
  \Delta_\kappa (x) \sim \frac{- 2 \kappa \, \Gamma (1 + i \kappa x^0)
  \cosh (\frac{\pi}{2} \kappa x^0 )}{(2\pi)^2\sqrt{1 + \frac{m^2}{\kappa^2} }
  \, |\overrightarrow{x} | (\kappa  |\overrightarrow{x} |)^{1 + i\kappa x^0} }\, .
\end{equation}

Further steps is to calculate the $\kappa$-deformed Feynman propagator and simple
Feynman diagrams, e.g. self-energy diagram
 in $\kappa$-deformed $\Phi^4$ theory. It has been already
 observed in \cite{koluma45}  that at the $\kappa$-deformed Feynman vertices the
 fourmomentum is not conserved, because the energy-momentum conservation is
 replaced as follows\footnote{Recently the nonconservation of four-momenta for
 Lie-algebraic noncommutative space-times has been rediscovered in
   \cite{43}.}

\begin{equation}\label{lul6.26}
  \delta \left(  \sum\limits_{n}^{i=1} p^{(i)}_\mu  \right)
  \longrightarrow \delta \left(\Delta^{(n)}_{\mu}
  (p^{(1)}_\mu , \ldots , p^{(n)}_\mu ) \right)\, ,
\end{equation}
where $\Delta^{(n)}_\mu$ is given by the formulae (5.15a-b).
 The renormalizaton of self-energy
 diagrams in $\kappa$-deformed
  $\Phi^4$ theory, in particular the problem of $UV/IR$ divergencies in such
 a framework, is now under consideration.

\sec
\section{Discussion}

In this paper we present the results related with so-called
$\kappa$-deformations of relativistic symmetries which
introduce the  elementary length $\lambda_\kappa =
\frac{\hbar}{\kappa c} $ as third fundamental constant besides
$c$ and $\hbar$. The appearance of this third universal constant
implies the existence of new  domain of ultrashort distances
$|x|\leq \lambda_\kappa$, where new noncommutative physics should
be applied. Calling this domain $\kappa$-relativistic physics we
obtain the following relation between the theories (see Table 1).
%%%%%%%%%
%%%%%%%%%%%%%%

\begin{table*}[h]
\begin{center}
\caption{}
\begin{tabular}{|c|p{105pt}|p{105pt}|p{105pt}|}
\hline
%\raisebox{0pt}[16pt][6pt]{ }
&
 $c=\infty ,\quad \kappa=\infty$
 &  $c$ finite, $\kappa=\infty$
 &{ $c$ finite, $\kappa$ finite
} \\[3pt]
\hline
%\raisebox{0pt}[16pt][6pt]
{}
& nonrelativistic
& relativistic
&
$\kappa$-relativistic
\\[6pt]
%\hline
%\raisebox{0pt}[16pt][6pt]
{$\hbar =0$}
&
 classical physics
 &
%\raisebox{-6pt}[16pt][6pt]
 {classical physics}
&
 classical physics
\\[6pt]
\hline
%\raisebox{0pt}[16pt][6pt]
{}
& nonrelativistic
&
relativistic
&
$\kappa$-relativistic
\\[6pt]
%\hline
%\raisebox{0pt}[16pt][6pt]
{$\hbar \neq 0$}
 &  quantum physics
&
{quantum physics}
&
quantum physics
\\[6pt]
\hline
\end{tabular}
\end{center}
\end{table*}

%%%%%%%%%
%%%%%%%%
%%%%%%%%
In this table  fourth pair of possibilities -
 $\kappa$-nonrelativistic physics - is not included because very
 short distances imply large velocities of test particles  probing
  the short-distance behaviour, so the nonrelativistic
  restriction of velocities
  is not reasonable for the distances $|x| <
  \lambda_\kappa$.\footnote{In fact the efforts to obtain the
   nontrivial deformation  of nonrelativistic physics by considering the limit
    $c\to \infty$ in the framework of  $\kappa$-deformed Poincar\'{e} symmetries
      were rather not successful.}

      The need for third fundamental constant has been also
      advocated by astrophysical considerations, where without
      Hopf-algebraic framework the simplest deformations of the
      mass-shall condition
\begin{equation}\label{lll7.1}
  p^2_0 = p^2 + m^2 \to p^2_0 = p^2 + m^2 + \alpha
  \frac{p^3}{\kappa}\, ,
\end{equation}
has been extensively studied (see for
      example \cite{koma35,koma47}).
 In (6.1)
 $\alpha$ is a dimensionless parameter and $\kappa$ can be
identified with the Planck mass ($\sim 10^{19}$Gev in energy
units).

At present the crucial question remains open
  how to incorporate full
Hopf algebra structure of $\kappa$-deformed symmetry algebras
into the phenomenological description of the ultra-short
 distance corrections implied by
quantum gravity. One of
  the points to be understood is the
 physical meaning of nonabelian symmetry of the quantum coproduct rules,
  in particular the problem of physically plausible description of quantum-deformed
   multiparticle states and energy-momentum conservation in
   $\kappa$-deformed field theory.
  These problems are
now under continuous considerations.

\subsection*{Acknowledgments}

One of the authors (J.L.) would like to thank J. Kowalski-Glikman for
 his interest in the subject, discussions and inducing us
  to finish this paper, which was written in its preliminary
  version in 1997--98.

%\frenchspacing
\def\PLB#1{Phys. Lett. {\bf B#1}}
\def\JPA#1{J. Phys.{\bf A#1}}

\end{document}